\documentclass{emulateapj}
\shorttitle{1046+59 325~MHz Transient Search}
\shortauthors{Jaeger et al.}

\usepackage{graphicx}
\usepackage{amssymb}
\usepackage{amsmath}
\usepackage{natbib}

\newcommand{\mjb}{mJy beam$^{-1}$}

\begin{document}

\title{Discovery of a Meter-Wavelength Radio Transient in the SWIRE Deep Field: 1046+59}

\author{T. R. Jaeger}
\affil{National Research Council Postdoctoral Research Associate, \\
	US Naval Research Laboratory, Code 7213, \\
	Washington, DC 20375 \\ 
	\it{ted.jaeger.ctr@nrl.navy.mill}}
\author{S. D. Hyman}
\affil{Department of Physics and Engineering, Sweet Briar College, \\
	Sweet Briar, Virginia 24595, USA}
\author{N. E. Kassim}
\affil{US Naval Research Laboratory, Code 7213, \\
	Washington, DC 20375}
\and
\author{T. J. W. Lazio}
\affil{Jet Propulsion Laboratory, California Institute of Technology \\
	Pasadena, CA 91106}

\begin{abstract}
We report the results of a low frequency radio variability and slow transient search using archival observations from the Very Long Array. We selected six 325~MHz radio observations from the spring of 2006, each centered on the Spitzer-Space-Telescope Wide-area Infrared Extragalactic Survey (SWIRE) Deep Field: 1046+59. Observations were spaced between 1~day to 3~months, with a typical single-epoch peak flux sensitivity below 0.2~\mjb near the field pointing center. We describe the observation parameters, data post-processing, and search methodology used to identify variable and transient emission. Our search revealed multiple variable sources and the presence of one, day-scale transient event with no apparent astronomical counterpart. This detection implies a transient rate of 1$\pm$1 event per 6.5 $\deg^2$ per 72 observing hours in the direction of 1046+59 and an isotropic transient surface density $\Sigma = 0.12 \deg^{-2}$ at 95\% confidence for sources with average peak flux density higher than 2.1~mJy over 12~hr.
\end{abstract}


\section{Introduction}
\label{sec:intro}
A long list of sources may present variable and transient behavior, from stellar flares \citep{Bastian1998ARAA, Osten2005ApJ}, spinning neutron stars \citep{Hewish1968Natur, Camilo2006Natur}, and gamma ray bursts \citep{Klebesadel1973ApJ, Dessenne1996MNRAS}, to yet-detected extra-solar planets \citep{Zarka1998JGR} and gravitational wave counterparts \citep{Blanchet2002LRR, Abbott2009PhRvD}. Radio observations serve an important role in the investigation of these objects and their variable behavior. Measurements at radio wavelengths uniquely probe magnetic field topology and non-thermal processes. This provides key information for understanding the acceleration mechanisms responsible for intense, time-variable emission. Past exploration of radio variability has been significantly limited in frequency range, time resolution, and spatial coverage. Recent improvements in low frequency radio spectrometer sensitivity, radio frequency interference (RFI) mitigation, and wide-field imaging techniques now place radio observations at the fore-front for the detection and subsequent understanding of these dynamic objects.

Early blind searches have already uncovered multiple variable and transient radio sources \citep{Hyman2005Natur, Lorimer2007Sci, Bower2007ApJ, Burke-Spolaor2010MNRAS, Bower2010ApJ}. However, a majority of the observations were performed at gigahertz (cm to mm wavelength) frequencies, leaving the sub-gigahertz (meter wavelength) sky vastly unexplored. We present the results from a sub-gigahertz variability and transient search using data from the Very Large Array (VLA) archive. We describe the observation and analysis of 6$\times$12~hr observations at 325~MHz (0.9~m, 6.5~$\deg^2$ field of view) with a common pointing center and temporal spacing between 1~day and 3~months. We report trends in the variability of approximately 950 field sources, highlighting multiple sources with significant flux variability. Further, we report the discovery of one transient event unassociated with a previously cataloged source and discuss the implications of our detection on existing transient limits. 

\section{Observations and Analysis}
\label{sec:obs}

\begin{table*}[t!]
\caption{Summary of radio observations listing the 2006 start date, observation time range (UTC), bandpass calibrator, amplitude and phase calibrator, number of UV visibilities after flagging, and RMS radio flux (\mjb) at the pointing center.}
\begin{center}
\begin{tabular}{|c|c|c|c|c|c|}
\hline
Start Date	   & Time Range 		& BP		     & A\&P Cal  	 & Visibilities & RMS  	\\
\hline\hline   
02/19	   & 02:15:59 - 14:13:39	& 0137+331  & 0542+498	 & 1660567   & 0.179 	\\
02/27	   & 01:45:20 - 13:42:19	& 0137+331  & 0137+331	 & 2076319   & 0.171 	\\
03/03	   & 01:30:29 - 03:26:29	& 0137+331  & 0542+498	 & 1622207   & 0.184 	\\
03/04	   & 01:25:49 - 13:22:40	& 0137+331  & 0542+498	 & 1379554   & 0.190 	\\
03/10	   & 01:02:59 - 12:58:59	& 0137+331  & 0542+498	 & 1725216   & 0.174 	\\
05/16	   & 20:36:09 - 08:31:34	& 0137+331  & 0542+498	 & 1395740   & 0.192 	\\
\hline
\end{tabular}
\end{center}
\label{tab:obs}
\end{table*}

\subsection{Radio Observations}
We analyzed six 325~MHz radio observations from the VLA archive centered on $\alpha=10^h46^m$, $\delta=+59^d$ recorded between February 19th, 2006 and May 17th, 2006 (see Tab. \ref{tab:obs}). The data was originally part of a larger effort by F. Owen to study the multi-wavelength properties of the general radio source population \citep{Owen2009AJ}. Individual field measurements have nearly identical observation parameters, providing an ideal dataset to search for source variability and transient emission.

Each observation lasted 12~hours and utilized two 6.25~MHz bands centered at 327.5~MHz and 321.6~MHz. Approximately 11~hours were allocated for measurements of the target field, with the remaining time allocated for calibrator observations. Data was collected in pseudo-continuum mode, with 15 spectral channels requested per band to mitigate radio frequency interference (RFI) and reduce bandwidth smearing of bright out-of-field confusing sources. The radio array consisted of 23, 25~m antennas in A-Configuration during each observation\footnote{Array re-configuration began before the final observation, however, only the sub-set of antennas and locations common to all epochs were chosen for our analysis.} (maximum baseline $\sim$ 35~km), resulting in an approximate 6" x 5" angular resolution and $\sim 6.5\deg^2$ field of view (based on a 2.9~$\deg$ full-width-half-power primary beam, see Sec. \ref{sec:id}).  

\subsection{Field Calibration and Imaging}
\label{sec:calib}
To ensure uniform processing of the radio data, we developed an automated reduction pipeline based on standard routines from the AIPS \footnote{Astronomical Image Processing System, release 31DEC10} and {\it Obit} \footnote{Obit is developed and maintained by Bill Cotton at The National Radio Astronomy Observatory in Charlottesville, Virginia, USA and is made available under the GNU General Public License. version 1.1.269-6-64b. } software packages. Each observation was processed as follows:
\begin{enumerate}
\item Initial data flagging was performed to eliminate two channels with known spectral RFI and all UV samples stronger than 50~Jy.
\item UV data was calibrated using standard AIPS tasks. The receiver bandpass correction was determined using calibration source 0137+331. The radio flux scale and antenna phase offsets were determined using a single 5 minute scan of 0542+498 (assumed 46.0~Jy at 327.5~MHz, 46.2~Jy at 321.6~MHz) for all epochs except February 27. For this epoch, all calibration was performed using 0137+331 (assumed 41.5~Jy,41.8~Jy) due to substantial RFI present during the 0542+498 scan.
\item  An additional antenna phase correction was performed by referencing a field source model based on the Westerbork Northern Sky Survey \citep[WENSS,][]{Rengelink1997AAS}.
\item Each field was imaged and CLEANed to a minimum flux level of 1~mJy. Field sources were subtracted to create a residual UV dataset. A second (and more extensive) round of RFI excision was performed by comparing the original UV data to the residual, using {\it Obit} tasks {\it LowFRFI} and {\it AutoFlag}.
\item The fully flagged UV data was re-imaged using {\it Obit} task {\it Imager} which is similar to AIPS IMAGR but can also execute self-calibration loops and auto-identify bright confusing sources. We preformed two rounds of phase-only self-calibration and an addition round of amplitude and phase self-calibration for each observation. Final RMS amplitude and phase variations were generally below 3\% and 0.5~$\deg$, respectively. Field images were made with a 3~$\deg$ diameter field-of-view (slightly larger than the primary beam FWHP) and 1$"$ pixels. Radio sources were automatically windowed and CLEANed to a level of 0.1~mJy.  
\end{enumerate}

\subsection{Source Identification and Filtering}
\label{sec:id}
Sources whose peak flux density exceeded 3 times the RMS flux density \footnote{Global $3\sigma$ before primary beam correction. We anticipate a large number of "accidental" detections at this threshold. A second threshold is applied later, based on Gaussian noise expectations.} were cataloged using {\it Obit} task {\it FndSou}. Individual catalogs were created for each epoch. Source characteristics were determined using a two-dimensional gaussian fit, providing an estimate of the center position, shape, peak and integrated flux density, local background RMS (within 300"), and amount of background slope or curvature. The catalogs were automatically filtered to exclude objects smaller than the synthesized beam. Highly extended sources (i.e. diameter $>$ 30") were not well modeled by a single 2D gaussian and were also excluded. All flux estimates were corrected for primary beam attenuation. 

\begin{figure}[t!]
\begin{center}
\includegraphics[width=3.05in]{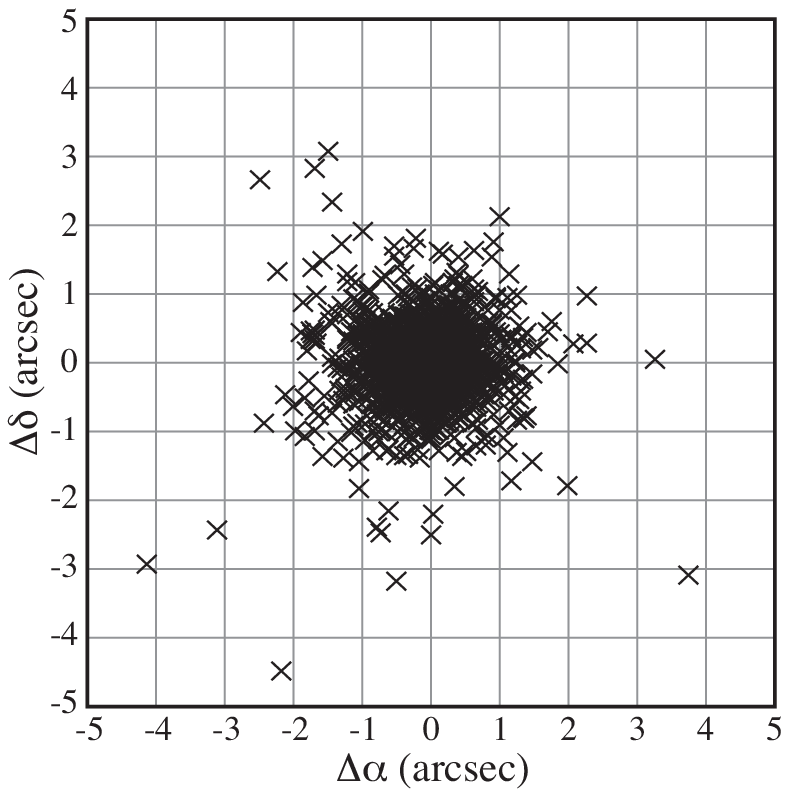}
\caption{Variation in the measured position ($\Delta\alpha, \Delta\delta$) of field sources identified in all observing epochs. Compact sources were generally located within 1$"$ with larger separations typically associated with inconsistent fitting of AGN.}    
\label{fig:pos}
\ \\
\ \\
\ \\
\includegraphics[width=3.05in]{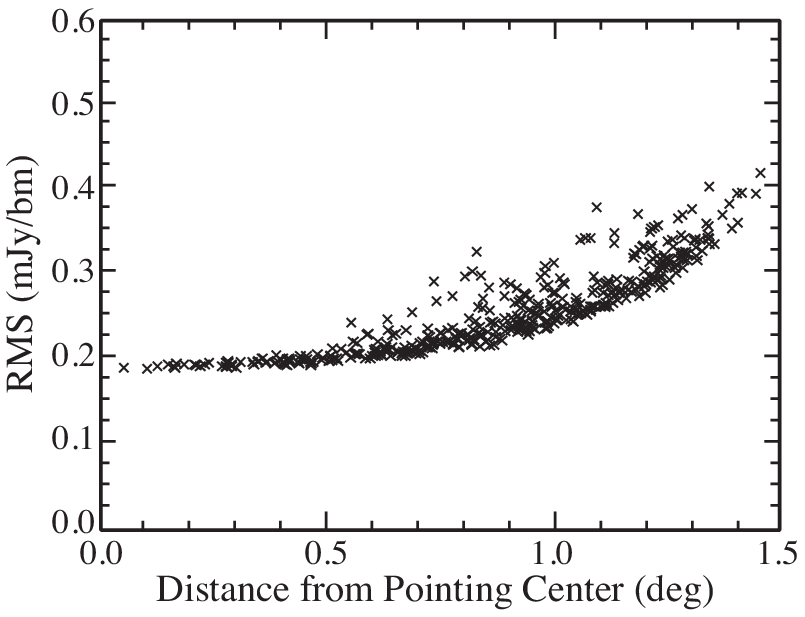}
\caption{Average RMS radio flux versus radial distance from the field center near sources detected in all epochs. Values are determined via histograms of the radio flux measurements within 300" of each source. The RMS measurements trace the attenuation of the primary beam, indicating a half-power radius of $\sim 1.45\deg$.}
\label{fig:rms}
\end{center}
\end{figure}

Sources from the six individual catalogs were matched across epoch by center position. A 5" matching radii (approximately one synthesized beam width) was used to account for fit error and/or any potential ionospheric refraction. Sources positioned within 5" were further discriminated using the peak flux density. The measured variation in source position ($\Delta\alpha, \Delta\delta$) between the mean position and the position observed for each epoch is plotted in Figure \ref{fig:pos}. Changes in source position were generally less than 1$"$ for all locations inside the primary beam. All displacements greater than 2" corresponded to inadequate automated fitting of extended sources with complex structure (see Sec. \ref{sec:var} for more discussion on the behavior of resolved and unresolved sources). Figure \ref{fig:rms} displays the average single epoch RMS ($1\sigma$) radio flux versus distance from the pointing center for sources persistent in all observations. The RMS measurements ranged from 0.19~\mjb to 0.41~\mjb and roughly trace primary beam attenuation profile. RMS values indicate a $\sim 1.45\deg$ half-power radius and no discernible beam asymmetry.   
 
The combined catalog was divided into portions representing (Group 1) sources detected in all 6 epochs, (Group 2) sources detected intermittently, i.e. sources found in multiple but not all epochs, and (Group 3) sources detected in only one epoch. To help identify detections in Group 2 and Group 3 which are associated with low signal-to-noise field sources (sources in a given epoch with peak flux below the initial global $3\sigma$ threshold), we created and cataloged a composite image made from the 6 individual observations. This image had a RMS $<$ 0.1~\mjb near the pointing center and contained over 1300 sources with integrated flux density exceeding 0.5~mJy. The number of sources and the integrated flux density values in our composite catalog were consistent with number and flux measurements of the same field made by \citet{Owen2009AJ}. Any single and intermittent detections coincident (5" search radii) with sources in the composite catalog were reevaluated to determine their properties for all 6 epochs. Approximately 90\% of the intermittent detections were reclassified and added to the list of persistent field sources (i.e. Group 1) during this process. No single epoch detections were reclassified during this process. All sources in Group 1 were searched for evidence of variability (see Sec. \ref{sec:varmes}). The remaining source detections from Group 2 and Group 3 were filtered by signal-to-noise ratio to identify statistically significant transient events. This filter is described in Section \ref{sec:trans_criteria}.    

\subsection{Measuring Source Variability}
\label{sec:varmes}
We searched the list of field sources identified in all observing epochs (Group 1) for variability in the measured total flux density. We quantify the flux variation using two methods, calculating the (1) modulation index $m$, i.e. standard deviation divided by the mean as used by \citet{Gaensler2000PASA}, and (2) a linear fit spanning the 6 epochs. Moreover, we checked all sources for any variation in shape. Changes in source shape ($\Delta \theta$) were quantified by first computing the standard deviation in the source major and minor axis from their mean value, normalizing each result by the mean value, and finally summing in quadrature. 

We assume that the flux density for a majority of the radio sources is stable over timescales longer than 3 months. However, source measurements from the individual epochs appeared uniformly offset from their mean value, presumably caused by long-duration amplitude changes in the primary amplitude calibrator. To account for this effect, we normalized each observation to the composite image catalog described in Section \ref{sec:id}. The gain corrections (one for each epoch) were determined by minimizing the standard deviation in total flux of the 100 brightest field sources from their composite value.  The resulting gain corrections ranged between 2\% (February 27) and 14\% (May 16). These corrections are consistent with VLA reported flux density variations for calibrator sources 0137+331 and 0542+498 during our observation period. Further, gain corrections did not appear spatially correlated. Performing the above normalization process for sub-regions inside the field of view produced similar results. 

\subsection{Transient Source Detection}
\label{sec:trans_criteria}

\begin{figure*}[t!]
\begin{center}
\includegraphics[width=6.04in]{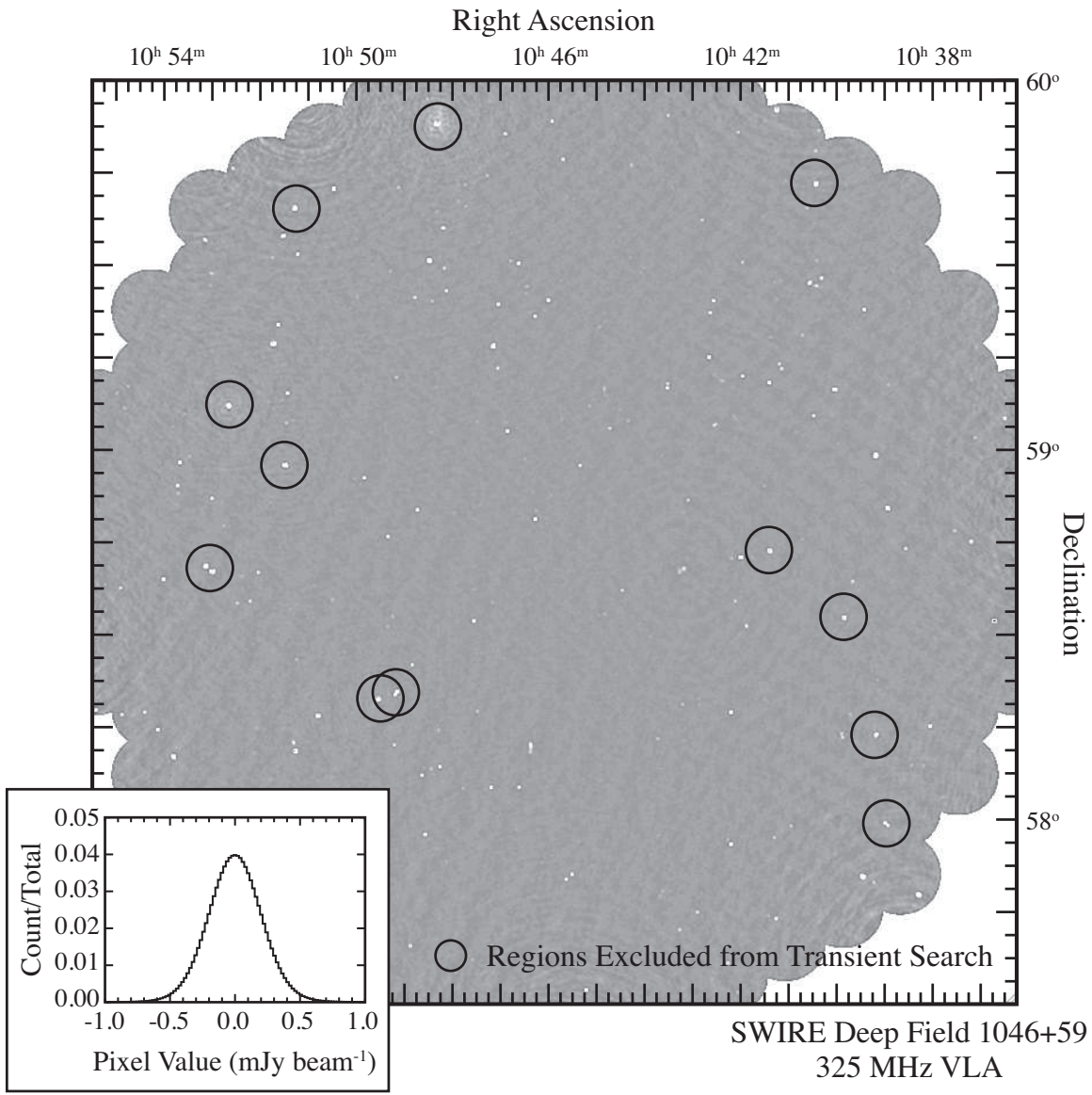}
\caption{(full) 325~MHz radio image centered on $\alpha=10^h46^m$, $\delta=+59^d$ observed February 19, 2006. The image has been corrected for primary beam attenuation. (inset) Flux density histogram. The image noise distribution is approximately Gaussian.}
\label{fig:field}
\end{center}
\end{figure*}

We searched the list of intermittent (Group 2) and single-epoch (Group 3) detections for previously undetected transient events. A vast majority of the "candidate" transients were due to noise fluctuations or systematic effects corresponding to bright field sources. Assuming the image noise can be modeled using Gaussian statistics (cf. Fig. \ref{fig:field} inset for verification), the expected number of transient detections $N$ above a given threshold $n_{\sigma}$ in a given field can be estimated as
\begin{equation}
N = n_{s} \cdot erfc(n_{\sigma}/\sqrt{2}),   
\label{eqn:nsig}
\end{equation}
where {\it erfc} is the complementary error function and $n_{s}$ is the number of independent samples 
\begin{equation}
n_{s} = \frac{n_{obs} \times \Omega_{obs}}{\Omega_{b}} \nonumber
\end{equation}
given by the number of epochs $n_{obs}$ times the observable area for each epoch $\Omega_{obs}$ divided by the area of the synthesized beam $\Omega_{b}$. Note that, in general, the number of independent samples $n_s$ is also dependent on the number of UV visibilities. For example, an image made from only the longest observation baseline will have no independent samples. However, as we are primarily focused on long duration transients and our VLA observations generate a large number of uniformly spaced visibilities per hour, we have omitted this dependency from our analysis. Such consideration may be required to analyze smaller timescales or when using other radio arrays. Further, the maximum observable area varies slightly with threshold choice due to diminishing beam sensitivity away from the pointing center (cf. Fig. \ref{fig:rms}). In practice, one chooses a flux threshold such that the predicted number of "accidental" detections is much less than one. For this observation, $n_{s}$ is approximately $2\times10^7$ using a search area which extends to the beam half-power points and $N$ is unity for a signal-to-noise threshold $n_{\sigma} \sim 5.5.$ We define the signal-to-noise as the ratio of the source fitted peak to the local RMS noise. Flux measurements larger than $5.5\sigma$ are then either due to real sources or indicate systematic effects.

Accidental detections associated with systematic effects were mitigated by excluding candidates within 10$'$ of source with median peak flux exceeding 100~\mjb. A grey-scale image of the 1046+59 field is shown in Figure \ref{fig:field}. The image has been corrected to reflect the primary beam attenuation profile. Circles mark the 12 sources brighter than 100~\mjb and the corresponding 10$'$ regions used to filter accidental detections caused by strong residual side-lobes. The excluded regions account for an approximate 5\% reduction in the transient search area $\Omega_{obs}$.  

\begin{figure}[t!]
\begin{center}
\includegraphics[width=3in]{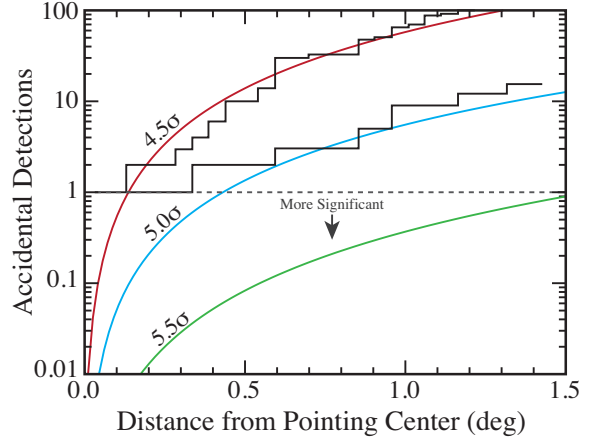}
\caption{Number of accidental detections versus distance from the pointing center. Histograms indicate measured values. Solid curves indicate detection estimates for different flux thresholds $n_{\sigma}$, assuming a Gaussian field noise distribution and incorporating the primary beam attenuation profile. A single transient detection is statistically significant only when the predicted number of 'accidental' detections at the same flux or SNR threshold is less than unity (dashed line).}    
\label{fig:tradeoff}
\end{center}
\end{figure}

Sensitivity variation across the primary antenna beam (cf. Fig. \ref{fig:rms}) manifests as a trade-off between the observing area $\Omega_{obs}$ and the desired detection threshold $n_{\sigma}$. Figure \ref{fig:tradeoff} illustrates the measured number of accidental detections versus distance from field pointing center. Curved lines indicate the expected number of accidental detections $N$ for different thresholds $n_{\sigma}$, assuming Gaussian field noise (cf. Fig. \ref{fig:field} inset for verification) and the measured primary beam attenuation profile. For this search, threshold of $5.5\sigma$ will detect significant events at any location inside the primary beam and maximize $\Omega_{obs}$. Detections exceeding $4.5\sigma$ are significant at all locations within 0.1~$\deg$ ($\sim$ 60-70 synthesized beams) of the pointing center. However, while the decreased threshold provides a factor of 2.5 better flux density sensitivity, the observing area for 4.5$\sigma$ events decreases by more than a factor of 200.

To identify statistically significant transient sources, we filtered the single-epoch detections to exclude sources with peak flux density below a 5.5$\sigma$, maximizing $\Omega_{obs}$. We filtered the intermittent detection list at a slightly lower threshold ($5\sigma$, equal to the minimum flux threshold for a single epoch) to provide more sensitivity to transients which occur between observations. To further evaluate the candidates, we required any positive transient event to be centered on a 300" diameter region free of spurious noise fluctuations at a 90\% expectation level, i.e. a background noise threshold of $4.1\sigma$ ($n_s \sim 2500$, $N \sim 0.1$ for $n_{\sigma}=4.1$). This is an important final step. Regions failing this criteria display non-Gaussian noise properties (likely due to systematic imaging effects) and the significance of any transient can not be determined by a single threshold test. Sources which remained after filtering were marked for visual inspection.         

\subsection{Transient Detection Rate}
The number of anticipated transient events $N_{t}$ is the product of the total observing area $n_{obs}\times\Omega_{obs}$ times the isotropic transient surface density $\Sigma$. Measurements at 843~MHz by \citet{Bannister2011MNRAS} suggest that $\Sigma$ may be as high as $1.3\times10^{-2}$~deg$^{-2}$ for long duration (day to year) transients stronger than 8~mJy. Assuming flat spectrum sources, we expect $\sim$ 1 transient detection from this search. However, there is still little known about the transient surface density frequency dependence and if scaling $\Sigma$ in such a manner is valid.       

The probability of detecting $N$ transients from an expected number of events $N_{t}$ is given as
\begin{equation}
P(N;N_{t})=\frac{N_{t}^N e^{-N_{t}}}{N!},
\label{eqn:poisson}
\end{equation}
assuming a Poisson distribution of sources. If $N_t$ is unknown (i.e. an unknown surface density), the measured number of transient detections can be used to establish an experimental upper limit for a set confidence level. In the limiting case of no detections ($N=0$), the expected number of events $N_{t}$ is approximately 3 at 95\% condidence ($P=0.05$) and the upper limit to the isotropic transient surface density can be expressed as
\begin{equation}
\Sigma < \frac{3.0}{n_{obs}\times\Omega_{obs}(n_{\sigma})}.
\end{equation}
$\Omega_{obs}(n_{\sigma})$ indicates that, in general, the observable area depends on the flux density threshold.


\section{Search Results}
\label{sec:results}

\subsection{Field Source Variability}
\label{sec:var}
Approximately 950 field sources were persistent in all six observations above an integrated flux density of $\sim$ 0.6~mJy (3$\sigma$ at the field center). Each of theses sources was searched for evidence of variability in the total flux density using the methods outlined in Section \ref{sec:varmes}. We find that the random variability in field sources sources is low and changes in a relatively uniform way.  Most detections have a modulation index on the order of 10\% for SNR $>$ 20 and below 5\% for SNR $>$ 50. The observed source variation is directly comparable to the fractional variations in the Gaussian major and minor axis fit parameters.

We observed excess variability in a small number of high SNR detections ($\sim$ 25 total sources, SNR $>$ 50). For a majority of these detection ($>$ 80\%), the measured modulation index appears associated with inadequate automated fitting of resolved sources with complex structure. In these specific cases, fit parameters (center position, major/minor axis, etc.) varied by as much as 50\% from the mean value and caused anomalous position and flux measurements. In the remaining cases, excess variability appears consistent with interstellar scintillation of a compact components in extragalactic radio sources. Meter-wavelength observations of $\sim$ 10~mas radio sources often display amplitude fluctuations ranging from 3\% to 10\% over timescales of days to months, respectively \citep{Blandford1986ApJ, Rickett1986ApJ}.

\begin{figure}[t!]
\begin{center}
\includegraphics[width=3.05in]{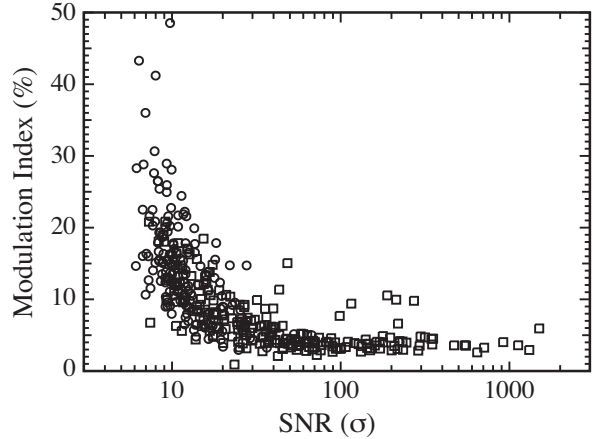}
\caption{Random variation in the 325~MHz radio flux density over 3 months vs. SNR ($\sigma$) for sources in the 1046+59 field. Unresolved sources are plotted with open circles, while resolved sources are plotted with open squares.}  
\label{fig:var}
\end{center}
\end{figure}

Table \ref{tab:var} lists sources which have a modulation index $m$ greater than that observed for field sources of similar SNR, but are not easily explained by variations in source shape $\Delta\theta$ caused by fit irregularities ($m \ge \Delta\theta$, see Sec. \ref{sec:varmes}). Redshift measurements from the NASA/IPAC Extragalactic Database (NED) are also given, if available. Figure \ref{fig:var} displays the 3-month flux variation (standard deviation divided by the mean) versus the signal-to-noise ratio (fitted peak divided by RMS) for all persistent field sources. Circles indicate measurements corresponding to unresolved sources, while squares indicate measurements corresponding to resolved sources. The brightest sources in the field are predominantly radio galaxies and are often resolved with meter-wavelength VLA observations. Sources with obvious fit irregularities have been excluded from Figure \ref{fig:var}, roughly characterized by shape variations greater than 15\% for sources with SNR $>$ 50.    

\begin{table*}[t!]
\caption{Sample variable sources in the 1046+59 field, listing the position ($\alpha,\delta$), Average RMS in \mjb, Average SNR, modulation index $m$ in percent, percent variation source shape $\Delta \theta$, and published redshift $z$ if known.}
\begin{center}
\begin{tabular}{|c|c|c|c|c|c|c|c|}
\hline
Criteria		& $\alpha$	& $\delta$		& RMS	& SNR	& $m$	& $\Delta \theta$	& $z$	\\
\hline\hline
			& 10 47 19.1   	& 58 21 17 	& 0.20 	&  213.30 	&   9.95 	&  5.30			& 1.22	\\
$m>7.5$		& 10 45 28.3   	& 59 13 27 	& 0.18 	&  188.90 	& 10.54 	&  5.63			& 2.31	\\
$SNR>50$	& 10 45 39.1   	& 58 07 11 	& 0.22 	&  115.51 	&   9.40 	&  5.27			& 1.15	\\
			& 10 44 50.5   	& 59 19 27 	& 0.18 	&   98.63	&   7.69 	&  5.84			& 0.84 	\\
\hline
			& 10 40 34.5   	& 59 54 45 	& 0.26 	&   48.31 	& 15.05 	&  3.89			& -   		\\
$m>10$		& 10 48 24.0   	& 58 30 27 	& 0.22 	&   38.46	& 10.15 	&  7.24			& -		\\
$SNR>20$	& 10 50 21.7   	& 58 42 57 	& 0.24 	&   27.68 	& 14.72 	&  6.97			& -		\\
			& 10 44 56.3   	& 59 38 02 	& 0.19 	&   21.26 	& 14.74 	&  6.07			& -		\\
\hline
\end{tabular}
\end{center}
\label{tab:var}
\end{table*}

In addition to measurements of random variability, we also monitored each persistent field source for linear trends, i.e. sources with significant brightening or dimming over the 3 month observation period. The two sources with the largest flux density trends are highlighted below.

\subsubsection*{J104719.1+582117}
J104719.1+582117 is optically identified as a quasar with an approximate distance of 350~Mpc \citep{Abazajian2009ApJS}. We observed a steady decrease in the flux density of J104719.1+582117 across the 6 observing epochs (see Fig. \ref{fig:varlc}, top). The measurements ranged from $47.7\pm0.39$~mJy on February 19th to $35.1\pm0.39$~mJy on May 17th, a 26\% decline in flux and a linear best-fit slope of -0.143~mJy/day.

\subsubsection*{J104539.1+580711}
\begin{figure}[t!]
\begin{center}
\includegraphics[width=3.05in]{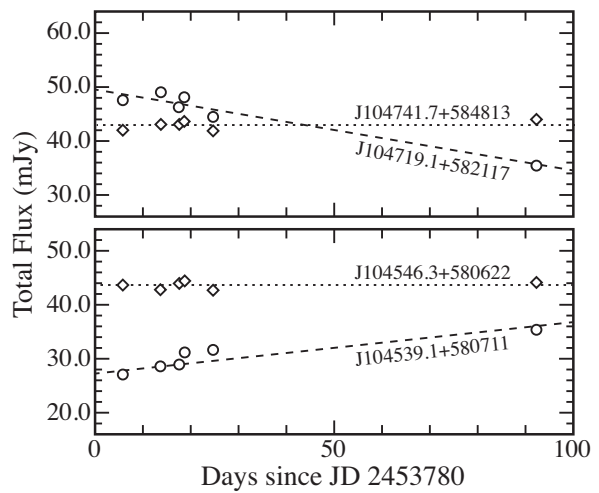}
\caption{Radio light curves for J104719.1+582117 (top) and J104539.1+580711 (bottom) from February 19, 2006 to May 17, 2006. Open circles indicate the integrated flux for the variable sources measured on each epoch. For reference, open diamonds indicate flux values for a nearby source of similar intensity. Dashed lines represent the best linear fit to the data. Uncertainty estimates are roughly equal to the symbol size.}  
\label{fig:varlc}
\end{center}
\end{figure}

J104539.1+580711 is a quasar located at approximately 330~Mpc \citep{Abazajian2009ApJS}. The peak radio flux density increased steadily over the observation epochs, growing from $27.1\pm0.42$~mJy on February 19th to $35.8\pm0.45$~mJy on May 17th. The best fit slope is 0.094~mJy/day (cf. Fig. \ref{fig:varlc}, bottom).

\ \\

\subsection{Transient Sources}
We inspected a list of 25 intermittent and 64 single-epoch transient candidates with peak flux density exceeding detection thresholds of $5\sigma$, $5.5\sigma$, respectively. All 25 intermittent detections and approximately 20 of the single-epoch detections were identified as weak persistent sources not recorded in our median catalog. Only one single-epoch source satisfied the final selection criteria given in Section \ref{sec:trans_criteria}, i.e. centered on 300" region free of noise fluctuations exceeding the 90\% expectation level. The remaining candidates were "false detections" which were marginally detected (44 of 46 below $5.7\sigma$) and in 300" regions plagued with imaging artifacts from a nearby field source. For completeness, we have excluded these regions from the final observation area ($\sim$ 1\% reduction to $\Omega_{obs}$), consistent with our mitigation of systematic effects around the 12 brightest field sources (cf. Sec. \ref{sec:id}). More details on the one remaining source are presented below.

\subsubsection*{J103916.2+585124}
\begin{figure}[t!]
\begin{center}
\includegraphics[width=3.05in]{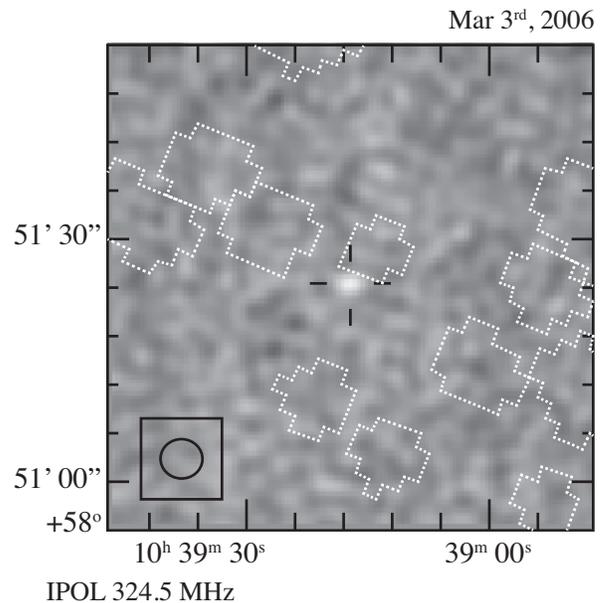}
\caption{Transient event J103916.2+585124 on March 3rd, 2006. The radio image scale is linear, ranging from -2~\mjb to 2~\mjb. The image RMS is 0.24~\mjb. White dotted lines outline the position of near-by galaxies measured by the Sloan Digital Sky Survey \citep[SDSS,][]{Abazajian2009ApJS}.}
\label{fig:T1039+58Gray}
\end{center}
\end{figure}

A transient radio source located at $\alpha=10^h39^m16.2^s$, $\delta=+58^d51^m24^s$ was detected in the March 3rd epoch with peak flux density equaling $1.70\pm0.25$~\mjb (6.7$\sigma$, 12~hr integration) and is centered on 300" region free of noise fluctuations in excess of 2.9$\sigma$. The source is unresolved and located 0.76~$\deg$ from the pointing center (15\% primary beam attenuation correction). J103916.2+585124 is positioned approximately 5" from an optical galaxy of undetermined type \citep{Abazajian2009ApJS}, but does not appear associated given the accuracy of the optical and radio position measurements (cf. Fig. \ref{fig:T1039+58Gray}). There are no published IR, X-ray, or GRB counterparts.

We reprocessed the March 3rd epoch in multiple ways to detect potential processing artifacts. We imaged the UV data isolating each frequency sub-band, each polarization, and without self-calibration. We also imaged the UV data with various pixel and facet sizes. The transient source was detected in all configurations. J103916.2+585124 appears unpolarized within the recorded error, measuring $1.76\pm0.32$~\mjb in right circular polarization (RR), $1.63\pm0.34$~\mjb in left circular polarization(LL), and is undetected at a 2.5$\sigma$ level of 0.68~\mjb in Stokes V . The source has no discernible frequency dispersion, with nearly equal intensity in each frequency sub-band ($1.67\pm0.31$~\mjb at 327.5~MHz, $1.72\pm0.31$~\mjb at 321.6~MHz). We did not detect J103916.2+585124 in images with frequency bandwidth less than 6.25~MHz. 

\begin{figure}[t!]
\begin{center}
\includegraphics[width=3.05in]{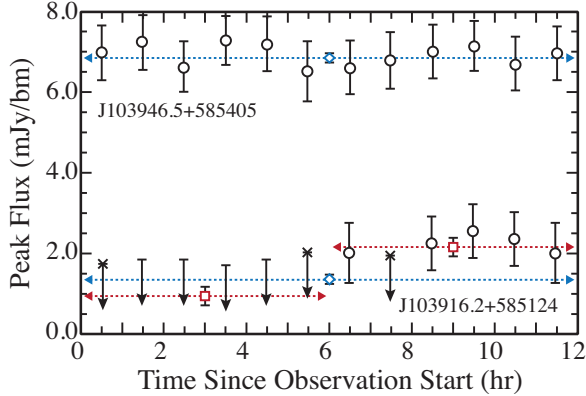}
\caption{Radio light curves for transient source J103916.2+585124 and a nearby field source on March 3, 2006. Open markers indicate the peak flux value at 12~hr (diamonds), 6~hr (squares), and 1~hr (circles) timescales with $1\sigma$ error bars. Visual detections with peak flux below 2.5$\sigma$ are indicated with crosses and downward arrows. The remaining non-detections are drawn with a 2.5$\sigma$ upper limit.}
\label{fig:T1039+58LC}
\end{center}
\end{figure}

A plot of the peak flux density versus time is shown in Figure \ref{fig:T1039+58LC} for J103916.2+585124 and a nearby field source J103946.5+585405. Markers indicate the peak value from a 2D Gaussian fit at 12~hr (diamonds), 6~hr (squares), and 1~hr (circles) intervals. We did not detect the source on timescales shorter than 1~hr, searching for emission on temporal spacings as fine as 10~s. Variability is evident on sub-day timescales, with the source appearing weak first 6 hours, then rising by approximately 1~\mjb to a maximum intensity of $2.1\pm0.29$~\mjb. The recorded signal-to-noise increases from 6.7 to 7.3 when only imaging the second half. J103916.2+585124 is undetected 4 days earlier on February 27th at a 2.5$\sigma$ level 0.60~\mjb and 12 hours later on March 4th at a 2.5$\sigma$ level 0.63~\mjb.

\subsubsection*{J103738.4+592001 and J104304.7+591726}
We note the presence of two events detected in the May 16th, 2006 epoch which have a peak flux exceeding 5.5$\sigma$ but appear in regions with non-Gaussian noise characteristics. J103738.4+592001 was observed with a peak flux of $1.74\pm0.29$~\mjb (6.1$\sigma$) and J104304.7+591726 was observed with a peak flux of $1.21\pm0.20$~\mjb (6.0$\sigma$). Similar to J103916.2+585124, each source is visually identified in 12~hr integration images which isolate each observing sub-band, Stokes R and Stokes L, and processing without self-calibration. Neither source has a contemporaneous optical, IR, X-Ray, or GRB counterpart. However, multiple noise fluctuations of comparable flux density ($> 5\sigma$, $< 4.1\sigma$ expected for 90\% false detection) are observed in close proximity to each source and the region noise could not be improved through reprocessing. Further, each source is undetected when dividing the observation into $2\times6$~hr intervals.   

\section{Discussion}
\label{sec:sd}
\begin{figure}[t!]
\begin{center}
\includegraphics[width=3.05in]{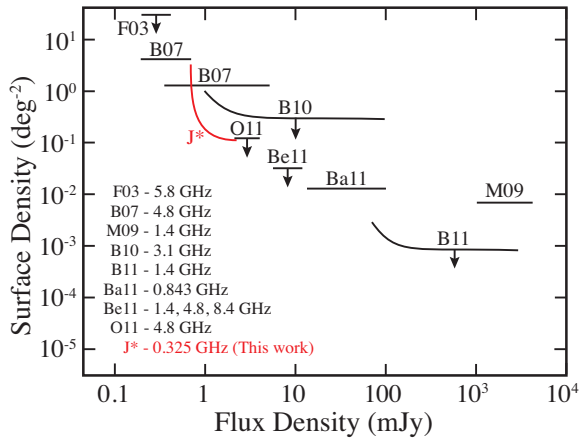}
\caption{Surface density versus radio flux density for transient searches with day-scale time resolution. Lines indicate results from \citet{Frail2003AJ} (F03), \citet{Bower2007ApJ, Bower2010ApJ, Bower2011ApJ} (B07, B10, B11), \citet{Matsumura2009AJ} (M09), \citet{Bannister2011MNRAS} (Ba11), and \citet{Bell2011MNRAS} (Be11). Arrows indicate 95\% confidence upper limits for searches resulting in no detections ($N=0$). Estimates based on this work (J*) are shown in red.}
\label{fig:SD}
\end{center}
\end{figure}

We detect one transient source exceeding a 5.5$\sigma$ minimum flux threshold (6.7$\sigma$ observed) located in a region free of spurious noise fluctuations. The measured transient rate in the direction of 1046+59 is then equal to 1$\pm$1 event per 6.5 $\deg^2$ per 72 observing hours for sources with average peak flux density higher than 2.1~mJy over 12~hr. We use Equation \ref{eqn:poisson} to report a corresponding isotropic surface density of $\Sigma = 0.12~\deg^{-2}$ 95\% confidence ($N=1, N_t = 4.5$). Figure \ref{fig:SD} displays the isotropic transient surface density inferred by this search along with results from past observations. Multiple transient searches have been published with time resolution ranging from sub-millisecond to multiyear. For a basis of comparison, we only depict searches which are sensitive to day to month temporal events. All published limits have been scaled to represent the Surface Density at 95\% confidence. The \citet{Bower2007ApJ} results have also been adjusted to reflect the recent reanalysis by \citet{Frail2011arXiv}. For completeness, we note the non-detection limit from \citet{Lazio2010bAJ} of $\Sigma < 10^{-7}$ with flux sensitivity $>$ 2.5~kJy at 74~MHz. This result has been omitted from Figure \ref{fig:SD}.

Our detection indicates a transient surface density that is more strict than the limit from \citet{Bower2007ApJ}, even considering the factor of 2 decrease suggested by \citet{Frail2011arXiv}. However, the frequency range in Figure \ref{fig:SD} (0.3~GHz to to 8.4~GHz) is greatly suppressed and this search was performed at frequency nearly and an order of magnitude lower lower than previous results with similar flux sensitivity. In general, gigahertz radio observations typically probe non-thermal synchrotron sources while sub-gigahertz observations are sensitive to plasma effects. Different searches may be sensitive to different source populations.

Transient J103916.2+585124 has no X-Ray or GRB counterpart, and is located approximately 5" from optical galaxy SDSS J103915.88+585127.7. The transient emission my be explained by interstellar scintillation of a compact radio source associated with SDSS J103915.88+585127.7 which is below the detection threshold. However, the source is positioned nearly one synthesized beam away from the galaxy center and the total flux doubles on a timescale as short as 6 hours (see Fig. \ref{fig:T1039+58LC}). Typical scintillation effects at 325~MHz only create about 3\% source variability on similar timescales. Therefore, we consider scintillation effects an unlikely source of the detected emission. 

\begin{figure}[t!]
\begin{center}
\includegraphics[width=3.05in]{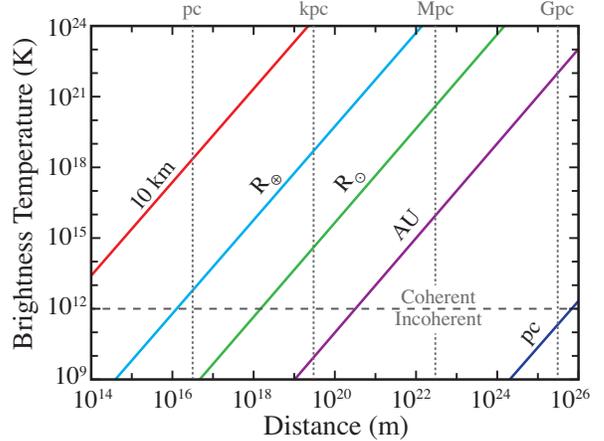}
\caption{Brightness temperature vs. distance estimates from Equation \ref{eqn:tb} considering a minimum radio detection level $S_{min} = 5.5 \times 0.38$~mJy over 12 hours. Diagonal lines represent values for sources of various radii. A horizontal dashed line at $10^{12}$~K denotes the approximate temperature boundary between coherent and incoherent emission processes.}
\label{fig:TB}
\end{center}
\end{figure}

To interpret the observation threshold for astrophysical sources implied by our transient detection, we express measured radio flux density $S$ in terms of a limiting source brightness temperature $T_B$. In the Rayleigh-Jeans limit, this can be written as
\begin{equation}
T_B = \frac{c^2 S}{2k_B\nu^2\Omega_s},   
\label{eqn:tb_full}
\end{equation}
where $\nu$ is the observing frequency and $\Omega_s$ is the source solid angle. Expressing the source solid angle in terms of a characteristic size $r$ and distance $D$, and inserting values relevant to this search, Equation \ref{eqn:tb_full} can be rewritten as  
\begin{equation}
T_B \sim 1\times10^8~\mathrm{K}\left(\frac{S}{\mathrm{mJy}}\right) \left(\frac{D}{\mathrm{pc}}\right)^2\left(\frac{R_{\sun}}{r}\right)^2,   
\label{eqn:tb}
\end{equation}
where $R_{\sun}$ is the solar radius. The range of astrophysical sources probed by our transient detection is illustrated in Figure \ref{fig:TB}.

Additional limitations to the transient progenitor can be inferred by considering the temporal characteristics illustrated in Figure \ref{fig:T1039+58LC}. We observe a $\sim$ 1~mJy increase in radio flux in 6 hours, implying an upper limit to the source size of $c\times6~hr \sim$ 42~AU. Using Equation \ref{eqn:tb}, incoherent emitters (approximate temperature boundary $1\times10^{12}$~K) of this size scale are detectable out to 900~kpc. 

We observe no evidence of variability on timescales less than one hour. This excludes sources with short temporal scales, such as pulsars and related phenomena \citep{Staelin1968Sci, Bhat2007ApJ}, CMI burst emission from ultra cool dwarfs \citep{Berger2001Nat, Osten2009ApJ}, and exoplanets \citep{Lazio2004ApJ}. Transient J103916.2+585124 also appears to decay on a time scale less than one day. This may be explained by the marginal detection of radio emission from supernovae. Spectral luminosity measurements by \citet{Weiler2002ARAnA} indicate that radio supernovae may be detected by this transient search at distances exceeding 20~Mpc.

We further consider the likelihood that our transient detection is due to coherent emission from a stellar flare. Observations of flare star EV Lac by \citet{Osten2005ApJ} revealed intense emission exceeding $10^{13}$~K with temporal structure similar to that observed for J103916.2+585124, i.e. minute-scale rise time and a decay time of hours. Further, flare stars are common in the solar vicinity, with a catalog of nearby UV Cet-type flare stars by \citet{Gershberg1999AAS} indicating 463 objects within 50 pc. Outbursts exceeding $2\times10^{14}$~K with physical size scales equal to one solar radii should be detected in our observations for sources within 1~kpc. While there is no published stellar counterpart, our astronomical detection distance is roughly twice the reported 2MASS non-detection limit for EV Lac type stars. 

\section{Summary}
\label{sec:summary}
We present the results from a 325~MHz variability and transient search using data from the VLA archive. The search revealed multiple variable sources well described by interstellar scintillation of extragalactic sources and two sources with significant linear trends. We also detect a single transient event at $\alpha=10h~39m~16.2s$, $\delta=+58d~51m~24s$ with flux density equaling 2.1~mJy ($6.7\sigma, 5.5\sigma$ "false detection" threshold). This discovery implies a transient rate of 1$\pm$1 event per 6.5 $\deg^2$ per 72 observing hours in the direction of 1046+59 and an isotropic transient surface density $\Sigma = 0.12 \deg^{-2}$ at 95\% confidence for sources with average peak flux density higher than 2.1~mJy over 12~hr.

\section*{Acknowledgements}
We would like to thank B. Cotton and W. Peters for assistance with Obit. SDH thanks F. Owen for sharing images previously made of this field for use in a preliminary analysis. Analysis utilizes data from VLA program AO201. The National Radio Astronomy Observatory is a facility of the National Science Foundation operated under cooperative agreement by Associated Universities, Inc. We also made use of the NASA/IPAC Extragalactic Database (NED) which is operated by the Jet Propulsion Laboratory, California Institute of Technology, under contract with the National Aeronautics and Space Administration. This research was performed while the primary author held a National Research Council Research Associateship Award a the US Naval Research Laboratory. Basic research in radio astronomy at the Naval Research Laboratory is supported by 6.1 base funding. Radio astronomy research at Sweet Briar College is funded by Research Corporation. Additional research was carried out at the Jet Propulsion Laboratory, California Institute of Technology, under a contract with the National Aeronautics and Space Administration. The {LUNAR consortium} is funded by the NASA Lunar Science Institute to investigate concepts for astrophysical observatories on the Moon. 

\bibliography{Master}

\begin{thebibliography}{36}
\expandafter\ifx\csname natexlab\endcsname\relax\def\natexlab#1{#1}\fi
\expandafter\ifx\csname url\endcsname\relax
  \def\url#1{\texttt{#1}}\fi
\expandafter\ifx\csname urlprefix\endcsname\relax\def\urlprefix{URL }\fi
\providecommand{\eprint}[2][]{\url{#2}}

\bibitem[{REV()}]{REVTEX41Control}


\bibitem[{08(1)}]{apsrev41Control}
08 1

\bibitem[{{Abazajian} et~al.(2009){Abazajian}, {Adelman-McCarthy},
  {Ag{\"u}eros}, {Allam}, {Allende Prieto}, {An}, {Anderson}, {Anderson},
  {Annis}, {Bahcall}, \& et~al.}]{Abazajian2009ApJS}
{Abazajian}, K.~N., {Adelman-McCarthy}, J.~K., {Ag{\"u}eros}, M.~A., {Allam},
  S.~S., {Allende Prieto}, C., {An}, D., {Anderson}, K.~S.~J., {Anderson},
  S.~F., {Annis}, J., {Bahcall}, N.~A., \& et~al. 2009, \apjs, 182, 543.
  \eprint{0812.0649}

\bibitem[{{Abbott} et~al.(2009){Abbott}, {Abbott}, {Adhikari}, {Ajith},
  {Allen}, {Allen}, {Amin}, {Anderson}, {Anderson}, {Arain}, \&
  et~al.}]{Abbott2009PhRvD}
{Abbott}, B.~P., {Abbott}, R., {Adhikari}, R., {Ajith}, P., {Allen}, B.,
  {Allen}, G., {Amin}, R.~S., {Anderson}, S.~B., {Anderson}, W.~G., {Arain},
  M.~A., \& et~al. 2009, \prd, 80, 102001. \eprint{0905.0020}

\bibitem[{{Bannister} et~al.(2011){Bannister}, {Murphy}, {Gaensler},
  {Hunstead}, \& {Chatterjee}}]{Bannister2011MNRAS}
{Bannister}, K.~W., {Murphy}, T., {Gaensler}, B.~M., {Hunstead}, R.~W., \&
  {Chatterjee}, S. 2011, \mnras, 412, 634. \eprint{1011.0003}

\bibitem[{{Bastian} et~al.(1998){Bastian}, {Benz}, \& {Gary}}]{Bastian1998ARAA}
{Bastian}, T.~S., {Benz}, A.~O., \& {Gary}, D.~E. 1998, \araa, 36, 131

\bibitem[{{Bell} et~al.(2011){Bell}, {Fender}, {Swinbank}, {Miller-Jones},
  {Law}, {Scheers}, {Spreeuw}, {Wise}, {Stappers}, {Wijers}, {Hessels}, \&
  {Masters}}]{Bell2011MNRAS}
{Bell}, M.~E., {Fender}, R.~P., {Swinbank}, J., {Miller-Jones}, J.~C.~A.,
  {Law}, C.~J., {Scheers}, B., {Spreeuw}, H., {Wise}, M.~W., {Stappers}, B.~W.,
  {Wijers}, R.~A.~M.~J., {Hessels}, J.~W.~T., \& {Masters}, J. 2011, \mnras,
  415, 2. \eprint{1103.0511}

\bibitem[{{Berger} et~al.(2001){Berger}, {Ball}, {Becker}, {Clarke}, {Frail},
  {Fukuda}, {Hoffman}, {Mellon}, {Momjian}, {Murphy}, {Teng}, {Woodruff},
  {Zauderer}, \& {Zavala}}]{Berger2001Nat}
{Berger}, E., {Ball}, S., {Becker}, K.~M., {Clarke}, M., {Frail}, D.~A.,
  {Fukuda}, T.~A., {Hoffman}, I.~M., {Mellon}, R., {Momjian}, E., {Murphy},
  N.~W., {Teng}, S.~H., {Woodruff}, T., {Zauderer}, B.~A., \& {Zavala}, R.~T.
  2001, \nat, 410, 338. \eprint{arXiv:astro-ph/0102301}

\bibitem[{{Bhat} et~al.(2007){Bhat}, {Wayth}, {Knight}, {Bowman}, {Oberoi},
  {Barnes}, {Briggs}, {Cappallo}, {Herne}, {Kocz}, {Lonsdale}, {Lynch},
  {Stansby}, {Stevens}, {Torr}, {Webster}, \& {Wyithe}}]{Bhat2007ApJ}
{Bhat}, N.~D.~R., {Wayth}, R.~B., {Knight}, H.~S., {Bowman}, J.~D., {Oberoi},
  D., {Barnes}, D.~G., {Briggs}, F.~H., {Cappallo}, R.~J., {Herne}, D., {Kocz},
  J., {Lonsdale}, C.~J., {Lynch}, M.~J., {Stansby}, B., {Stevens}, J., {Torr},
  G., {Webster}, R.~L., \& {Wyithe}, J.~S.~B. 2007, \apj, 665, 618.
  \eprint{0705.0404}

\bibitem[{{Blanchet}(2002)}]{Blanchet2002LRR}
{Blanchet}, L. 2002, Living Reviews in Relativity, 5, 3.
  \eprint{arXiv:gr-qc/0202016}

\bibitem[{{Blandford} et~al.(1986){Blandford}, {Narayan}, \&
  {Romani}}]{Blandford1986ApJ}
{Blandford}, R., {Narayan}, R., \& {Romani}, R.~W. 1986, \apjl, 301, L53

\bibitem[{{Bower} et~al.(2010){Bower}, {Croft}, {Keating}, {Whysong},
  {Ackermann}, {Atkinson}, {Backer}, {Backus}, {Barott}, {Bauermeister},
  {Blitz}, {Bock}, {Bradford}, {Cheng}, {Cork}, {Davis}, {DeBoer}, {Dexter},
  {Dreher}, {Engargiola}, {Fields}, {Fleming}, {Forster}, {Gutierrez-Kraybill},
  {Harp}, {Heiles}, {Helfer}, {Hull}, {Jordan}, {Jorgensen}, {Kilsdonk}, {Law},
  {van Leeuwen}, {Lugten}, {MacMahon}, {McMahon}, {Milgrome}, {Pierson},
  {Randall}, {Ross}, {Shostak}, {Siemion}, {Smolek}, {Tarter}, {Thornton},
  {Urry}, {Vitouchkine}, {Wadefalk}, {Weinreb}, {Welch}, {Werthimer},
  {Williams}, \& {Wright}}]{Bower2010ApJ}
{Bower}, G.~C., {Croft}, S., {Keating}, G., {Whysong}, D., {Ackermann}, R.,
  {Atkinson}, S., {Backer}, D., {Backus}, P., {Barott}, B., {Bauermeister}, A.,
  {Blitz}, L., {Bock}, D., {Bradford}, T., {Cheng}, C., {Cork}, C., {Davis},
  M., {DeBoer}, D., {Dexter}, M., {Dreher}, J., {Engargiola}, G., {Fields}, E.,
  {Fleming}, M., {Forster}, R.~J., {Gutierrez-Kraybill}, C., {Harp}, G.~R.,
  {Heiles}, C., {Helfer}, T., {Hull}, C., {Jordan}, J., {Jorgensen}, S.,
  {Kilsdonk}, T., {Law}, C., {van Leeuwen}, J., {Lugten}, J., {MacMahon}, D.,
  {McMahon}, P., {Milgrome}, O., {Pierson}, T., {Randall}, K., {Ross}, J.,
  {Shostak}, S., {Siemion}, A., {Smolek}, K., {Tarter}, J., {Thornton}, D.,
  {Urry}, L., {Vitouchkine}, A., {Wadefalk}, N., {Weinreb}, S., {Welch}, J.,
  {Werthimer}, D., {Williams}, P.~K.~G., \& {Wright}, M. 2010, \apj, 725, 1792.
  \eprint{1009.4443}

\bibitem[{{Bower} \& {Saul}(2011)}]{Bower2011ApJ}
{Bower}, G.~C., \& {Saul}, D. 2011, \apjl, 728, L14+. \eprint{1101.0121}

\bibitem[{{Bower} et~al.(2007){Bower}, {Saul}, {Bloom}, {Bolatto},
  {Filippenko}, {Foley}, \& {Perley}}]{Bower2007ApJ}
{Bower}, G.~C., {Saul}, D., {Bloom}, J.~S., {Bolatto}, A., {Filippenko}, A.~V.,
  {Foley}, R.~J., \& {Perley}, D. 2007, \apj, 666, 346. \eprint{0705.3158}

\bibitem[{{Burke-Spolaor} \& {Bailes}(2010)}]{Burke-Spolaor2010MNRAS}
{Burke-Spolaor}, S., \& {Bailes}, M. 2010, \mnras, 402, 855. \eprint{0911.1790}

\bibitem[{{Camilo} et~al.(2006){Camilo}, {Ransom}, {Halpern}, {Reynolds},
  {Helfand}, {Zimmerman}, \& {Sarkissian}}]{Camilo2006Natur}
{Camilo}, F., {Ransom}, S.~M., {Halpern}, J.~P., {Reynolds}, J., {Helfand},
  D.~J., {Zimmerman}, N., \& {Sarkissian}, J. 2006, \nat, 442, 892.
  \eprint{arXiv:astro-ph/0605429}

\bibitem[{{Dessenne} et~al.(1996){Dessenne}, {Green}, {Warner}, {Titterington},
  {Waldram}, {Barthelmy}, {Butterworth}, {Cline}, {Gehrels}, {Palmer},
  {Fishman}, {Kouveliotou}, \& {Meegan}}]{Dessenne1996MNRAS}
{Dessenne}, C.~A.-C., {Green}, D.~A., {Warner}, P.~J., {Titterington}, D.~J.,
  {Waldram}, E.~M., {Barthelmy}, S.~D., {Butterworth}, P.~S., {Cline}, T.~L.,
  {Gehrels}, N., {Palmer}, D.~M., {Fishman}, G.~J., {Kouveliotou}, C., \&
  {Meegan}, C.~A. 1996, \mnras, 281, 977

\bibitem[{{Frail} et~al.(2003){Frail}, {Kulkarni}, {Berger}, \&
  {Wieringa}}]{Frail2003AJ}
{Frail}, D.~A., {Kulkarni}, S.~R., {Berger}, E., \& {Wieringa}, M.~H. 2003,
  \aj, 125, 2299

\bibitem[{{Frail} et~al.(2011){Frail}, {Kulkarni}, {Ofek}, {Bower}, \&
  {Nakar}}]{Frail2011arXiv}
{Frail}, D.~A., {Kulkarni}, S.~R., {Ofek}, E.~O., {Bower}, G.~C., \& {Nakar},
  E. 2011, ArXiv e-prints. \eprint{1111.0007}

\bibitem[{{Gaensler} \& {Hunstead}(2000)}]{Gaensler2000PASA}
{Gaensler}, B.~M., \& {Hunstead}, R.~W. 2000, Publications Astronomical Society
  of Australia, 17, 72. \eprint{arXiv:astro-ph/9911194}

\bibitem[{{Gershberg} et~al.(1999){Gershberg}, {Katsova}, {Lovkaya},
  {Terebizh}, \& {Shakhovskaya}}]{Gershberg1999AAS}
{Gershberg}, R.~E., {Katsova}, M.~M., {Lovkaya}, M.~N., {Terebizh}, A.~V., \&
  {Shakhovskaya}, N.~I. 1999, \aaps, 139, 555

\bibitem[{{Hewish} et~al.(1968){Hewish}, {Bell}, {Pilkington}, {Scott}, \&
  {Collins}}]{Hewish1968Natur}
{Hewish}, A., {Bell}, S.~J., {Pilkington}, J.~D.~H., {Scott}, P.~F., \&
  {Collins}, R.~A. 1968, \nat, 217, 709

\bibitem[{{Hyman} et~al.(2005){Hyman}, {Lazio}, {Kassim}, {Ray}, {Markwardt},
  \& {Yusef-Zadeh}}]{Hyman2005Natur}
{Hyman}, S.~D., {Lazio}, T.~J.~W., {Kassim}, N.~E., {Ray}, P.~S., {Markwardt},
  C.~B., \& {Yusef-Zadeh}, F. 2005, \nat, 434, 50.
  \eprint{arXiv:astro-ph/0503052}

\bibitem[{{Klebesadel} et~al.(1973){Klebesadel}, {Strong}, \&
  {Olson}}]{Klebesadel1973ApJ}
{Klebesadel}, R.~W., {Strong}, I.~B., \& {Olson}, R.~A. 1973, \apjl, 182, L85+

\bibitem[{{Lazio} et~al.(2004){Lazio}, {Farrell}, {Dietrick}, {Greenlees},
  {Hogan}, {Jones}, \& {Hennig}}]{Lazio2004ApJ}
{Lazio}, T.~J., W., {Farrell}, W.~M., {Dietrick}, J., {Greenlees}, E., {Hogan},
  E., {Jones}, C., \& {Hennig}, L.~A. 2004, \apj, 612, 511

\bibitem[{{Lazio} et~al.(2010){Lazio}, {Clarke}, {Lane}, {Gross}, {Kassim},
  {Ray}, {Wood}, {York}, {Kerkhoff}, {Hicks}, {Polisensky}, {Stewart},
  {Paravastu Dalal}, {Cohen}, \& {Erickson}}]{Lazio2010bAJ}
{Lazio}, T.~J.~W., {Clarke}, T.~E., {Lane}, W.~M., {Gross}, C., {Kassim},
  N.~E., {Ray}, P.~S., {Wood}, D., {York}, J.~A., {Kerkhoff}, A., {Hicks}, B.,
  {Polisensky}, E., {Stewart}, K., {Paravastu Dalal}, N., {Cohen}, A.~S., \&
  {Erickson}, W.~C. 2010, \aj, 140, 1995. \eprint{1010.5893}

\bibitem[{{Lorimer} et~al.(2007){Lorimer}, {Bailes}, {McLaughlin}, {Narkevic},
  \& {Crawford}}]{Lorimer2007Sci}
{Lorimer}, D.~R., {Bailes}, M., {McLaughlin}, M.~A., {Narkevic}, D.~J., \&
  {Crawford}, F. 2007, Science, 318, 777. \eprint{0709.4301}

\bibitem[{{Matsumura} et~al.(2009){Matsumura}, {Niinuma}, {Kuniyoshi},
  {Takefuji}, {Asuma}, {Daishido}, {Kida}, {Tanaka}, {Aoki}, {Ishikawa},
  {Hirano}, \& {Nakagawa}}]{Matsumura2009AJ}
{Matsumura}, N., {Niinuma}, K., {Kuniyoshi}, M., {Takefuji}, K., {Asuma}, K.,
  {Daishido}, T., {Kida}, S., {Tanaka}, T., {Aoki}, T., {Ishikawa}, S.,
  {Hirano}, K., \& {Nakagawa}, S. 2009, \aj, 138, 787

\bibitem[{{Osten} et~al.(2005){Osten}, {Hawley}, {Allred}, {Johns-Krull}, \&
  {Roark}}]{Osten2005ApJ}
{Osten}, R.~A., {Hawley}, S.~L., {Allred}, J.~C., {Johns-Krull}, C.~M., \&
  {Roark}, C. 2005, \apj, 621, 398. \eprint{arXiv:astro-ph/0411236}

\bibitem[{{Osten} et~al.(2009){Osten}, {Phan-Bao}, {Hawley}, {Reid}, \&
  {Ojha}}]{Osten2009ApJ}
{Osten}, R.~A., {Phan-Bao}, N., {Hawley}, S.~L., {Reid}, I.~N., \& {Ojha}, R.
  2009, \apj, 700, 1750. \eprint{0905.4197}

\bibitem[{{Owen} et~al.(2009){Owen}, {Morrison}, {Klimek}, \&
  {Greisen}}]{Owen2009AJ}
{Owen}, F.~N., {Morrison}, G.~E., {Klimek}, M.~D., \& {Greisen}, E.~W. 2009,
  \aj, 137, 4846. \eprint{0904.2011}

\bibitem[{{Rengelink} et~al.(1997){Rengelink}, {Tang}, {de Bruyn}, {Miley},
  {Bremer}, {Roettgering}, \& {Bremer}}]{Rengelink1997AAS}
{Rengelink}, R.~B., {Tang}, Y., {de Bruyn}, A.~G., {Miley}, G.~K., {Bremer},
  M.~N., {Roettgering}, H.~J.~A., \& {Bremer}, M.~A.~R. 1997, \aaps, 124, 259

\bibitem[{{Rickett}(1986)}]{Rickett1986ApJ}
{Rickett}, B.~J. 1986, \apj, 307, 564

\bibitem[{{Staelin} \& {Reifenstein}(1968)}]{Staelin1968Sci}
{Staelin}, D.~H., \& {Reifenstein}, E.~C., III 1968, Science, 162, 1481

\bibitem[{{Weiler} et~al.(2002){Weiler}, {Panagia}, {Montes}, \&
  {Sramek}}]{Weiler2002ARAnA}
{Weiler}, K.~W., {Panagia}, N., {Montes}, M.~J., \& {Sramek}, R.~A. 2002,
  \araa, 40, 387

\bibitem[{{Zarka}(1998)}]{Zarka1998JGR}
{Zarka}, P. 1998, \jgr, 103, 20159

\end{thebibliography}
\end{document}